\documentclass[A4,twoside,12pt,reqno]{article}

\usepackage[super,compress]{cite}
\usepackage{graphicx}
\usepackage{float}
\usepackage[latin1]{inputenc}

\usepackage{amssymb}
\usepackage{amsthm}
\usepackage[tbtags]{amsmath}
\usepackage{doc}
\usepackage{latexsym}
\usepackage{amscd}
\usepackage[frame,cmtip,arrow,matrix,line,graph,curve]{xy}
\usepackage{graphics}
\usepackage{epsfig}
\usepackage{eucal}
\usepackage{mathrsfs}
\usepackage[margin=1.3in]{geometry}

\begin{document}
\thispagestyle{plain}
 \markboth{}{}
\small{\addtocounter{page}{0} \pagestyle{plain}

\noindent\centerline{\Large \bf Whether Lyra's Manifold itself is a Hidden Source} \\ 
\noindent\centerline{\Large \bf of Dark Energy }\\

\noindent\centerline{$^1$Kangujam Priyokumar Singh}\\
\noindent\centerline{$^2$Koijam Manihar  Singh}\\
\noindent\centerline{$^3$Mahbubur Rahman Mollah}\\
\textbf{}
\noindent\centerline{$^1$Department of Mathematical Sciences, Bodoland University,}\\
\noindent\centerline{Kokrajhar, Assam-783370, India. }\\
\noindent\centerline{pk\_mathematics@yahoo.co.in, Tel.: +919856134748}
\noindent\centerline{$^2$Department of Mathematics, ICFAI University,}\\
\noindent\centerline{Tripura -799210, India. }\\
\noindent\centerline{drmanihar@rediffmail.com, Tel.: +918974000569 }\\
\noindent\centerline{$^3$Department of Mathematics, Commerce College,}\\
\noindent\centerline{Kokrajhar, Assam - 783370, India }\\
\noindent\centerline{mr.mollah123@gmail.com, Tel.: +919854463450}\\

\footnote{Preprint of an article published in [Int. J. Theor. Phys. (2017) 56 : 2607 - 2621, DOI 10.1007/s10773-017-3417-4][Journal URL: www.springer.com/physics/journal/10773]}

\textbf{Abstract:} In the course of investigation of some interesting cosmic string universes in the five dimensional Lyra manifold it is excitingly found that the geometry itself of Lyra manifold behaves as a new source of dark energy and this energy takes a form similar to that of quintessence in most of the cases, though in one case the dark energy comes out to be that of the cosmological constant type. The behaviour of the universes and their contribution to the process of evolution are examined. Further study of such type of universes will be very helpful in explaining the present accelerated expansion behaviour of the universe.

\section{Introduction}           
\label{sect:intro}

We have been established and belived that our universe is accelerating at the present epoch instead of showing down as predicted by the Big Bang theory Silk [1]. However, till today we are not in a state to provide an exactly clear statement about the origin and evolution of our universe. When we study different literatures and philosophical point of views in this regard we found that different mines provide different opinions about our universe. Actually the universe is full of mysterious elements and numerous effects of interactions which cannot be detected and difficult to explain even with advanced technology. So the most challenging problems in Astrophysics and modem cosmology is to understand the late time acceleration of the universe. In recent years many researchers and Scientists are putting huge effort to explain the dynamics of the universe and to understand the future evolution of the universe with the attention in the context of dark energy and modified theories of gravity. On the other hand many prominent results of the cosmological observations like Type SNeIa supernovae [2-7], CMB (Cosmic Microwave Background) anisotropies [8-10], the large scale galaxies structures of universe [11-12], BAO (Baryon Acoustic Oscillations) [13-14], WMAP (Wilkinson Microwave Anisotropy Probe) [15-16] Sachs-Wolfe effects [17] and SDSS (Sloan digital sky survey) [18-21] are noticed to us for the cosmic acceleration with direct and indirect evidence. Not only the above mentioned observations and surveys, but some other new cosmological results and data sets like Planck [22-23], Atacama Cosmology Telescope (ACT) [24], South Pole Telescope Sunyaev-Zel'dovice (SPT-SZ) survey [25] have measured the temperature and polarization of  CMB to exquisite precision, are also supporting this fact. These results are more acceptable to the community and beneficial to the researchers to understand about the universe in this modern era.\\
\\
From the recent literatures and findings we know that dark energy dominates the universe with positive energy density and negative pressure, responsible to produce sufficient acceleration in late time evolution of the Universe. So dark energy becomes as a major outstanding issue in physics and cosmology today. There are a number of useful reviews of dark energy that are focus on theory [26-31], on probes of dark energy [32] and on the cosmological constant [33-34]. Also,many prominent researchers like Yoo and Watanabe [35], Bahrehbakhsh et al. [36] , Joyce et al. [37], Pasechnik [38], Wang et al. [39], El-Nabulsi [40-48] , Motloch et al. [49], Silbergleit [50], ; Linder [51], Josset et al. [52], Nojiri et al. [53], Nojiri and Odintsov [54-55], Felice and Tsujikawa [56] , Faraoni [57], Boehmer et al.[58], Tsujikawa [59] discussed different models about the dark energy in different context with the comparisons of observational findings. Some of the important claimants of dark energy are tachyons [60], chaplygin gas [61], phantom [62], k-essence and quintessence [63-65] along with other four elements i.e. dark matter, baryons, radiation and neutrinos. But so far there is no direct detection of such exotic fluids. Although the literature is now flooded with hundreds of model for dark energy what we lack is precise cosmological data coming from variety of observations involving both background and inhomogeneous universe that can discriminate among these models. In this connection if we accept that Einstein was correct with his general relativity theory to explain accelerated expansion of the universe could also be explained by negative pressure working against gravity. The belief of Einstein to the static universe made him to think about negative pressure which will stop the attraction of the gravity. However, we know that we have non static universe, moreover we know that we have accelerated expansion. According to the above observational data analysis, it can be estimated the amount of the negative pressure in our universe, which we call it as dark energy. The simple question about the nature of the dark energy is still one of the intriguing questions and left free space for new speculations.\\
\\
 Many Physicists have been investigating about gravitation in different contexts after Einstein. Hermann Weyl [66] attempted to generalize the idea of geometrizing the gravitation and electromagnetism by applying different techniques and methods. He described both gravitation and electromagnetism geometrically by formulating a new kind of gauge theory involving metric tensor with an intrinsic geometrical significance.  With the concept of Einstein's general theory of relativity  Lyra [67] suggested a modification  by introducing a gauge function into the structure less manifold which removes the non-integrability condition of the length of a vector under parallel transport. Such theories are commonly known as the modified theories of the gravitation or alternate theory of gravitation. Some important modified theories of gravitation are Brans-Dicke theory [68], Scalar-tensor theories[69], Vector-tensor theory [70], Weyl's theory [71], F(R) Gravity [72-76], Mimetic Gravity [77], Mimetic F(R)gravity [78,79], Lyra geometry [80] and many more. We can explain about the accelerating expansion of the universe in the context of these modified theories of gravitation. Out of these modified theories of gravitation, here we will discuss about the Lyra geometry.\\
\\
It is common to us that Lyra geometry is a modification of Riemanian geometry by introducing a gauge function into the structure less manifold which removes the non-integrability condition of the length of a vector under parallel transport. Lyra geometry along with constant gauge vector  $\phi_i$ will either play the role of cosmological constant or creation field ( equal to Hoyle's creation field [81-83]) which is discussed by Soleng [84]. Many researchers proposed different cosmological models in different context of Lyra geometry.  Sen [85] formulated a new scalar-tensor theory of gravitation based on Lyra geometry in which he found that static model with finite density in Lyra geometry is similar to the  Einstein's static model, but it exhibited red shift which is a significant difference of the model. Later, Sen and Dunn [86] ; Rosen [87]), suggested that this theory was based on non-integrability of length transfer so that it had some unsatisfactory features and hence this theory did not gain general acceptance. Halford [88, 89] pointed out that in the normal general relativistic treatment the constant displacement vector field $\phi_{k}$ in Lyra's geometry plays the role of cosmological constant. Also, as in the Einstein's theory of relativity the scalar-tensor treatment based on Lyra's geometry predicts the same effect, within observational limits, as far as the classical solar system test are concerned . Many authors like Rahaman et al. [90, 91], Casana et al. [92, 93] ; Mohanty et al. [94, 95] ; Mahanta et al. [96]; Asgar et al. [97]; Mollah et al.[98]; Mollah and Priyokumar [99] attempted to solve Einstein's field equations in the framework of Lyra's geometry and successfully find their solutions under different circumstances.\\
\\
  It is also learn that  the space-time symmetry plays a vital role in the features of space-time that can be described as exhibiting some form of symmetry. The importance of symmetries in relativity and cosmology is to simplify Einstein's field equations and to provide a classification of the space time according to the structure of the corresponding Lie algebra. Some authors like Henriksen et al. [100], Mohanty ang Samanta [101] , Rao and Neelima [102], Singh et al. [103], Cahill and Taub [104]) solved Einstein's field equations for an axially symmetric space-time by imposing certain conditions upon the scale factor of the space time together with some conditions on matter which represents such space-time and some restrictions on its physical properties. To explain the accelerated expansion of the universe many prominent researchers like Shchigolev [105] , Hova [106], Ali and Rahman [107], Megied et al. [108], Khurshudyan  et al. [109, -110], Saadat [111], Darabi et al. [112], Ziaie  et al. [113], Pucheu  et al. [114]) have investigated and proposed different cosmological models and ideas of the universe within the framework of Lyra's geometry and other theories of relativity in different context.\\
\\
 Earlier majority of  researchers in the field of Astrophysics, Cosmology and Particle Physics were doing their research in this area by considering only 4-dimensional case. But in this modern era peoples are more interested to study higher dimensional case, since the solutions of Einstein field equations in higher dimensional space times are believed to have physical relevance possibly at extremely early times before the universe underwent the compactification transitions. So to fit the observational data in a better way and to study the accelerated expansion of our universe many researchers like Bahrehbakhsh et al. [115], Rahaman [116], Chatterjee [117],  Chatterjee and Bhui [118], Chatterjee and Banerjee [119],Sil and Chatterjee [120], Rahaman et al.[121], El-Nabulsi [122-127], Rador [128], Kahil and Harko [129] discussed about different cosmological models concerning extra dimension. From the critical study we know that by using a suitable scalar field we can show that the phase transitions on the early universe can give rise to such objects which are nothing but the topological knots in the vacuum expectation value of the scalar field and most of their energy is concentrated in a small region.  So it is necessary for us to study the cosmological problems by considering higher-dimensional space-time  to unify gravity with other interactions. In cosmology, particularly for study of the early stage of universe the present four dimensional stage of the universe might have been preceded by a multi-dimensional stage. Recently, Priyokumar and Mollah [130] studied about higher dimensional LRS Bianchi type-I cosmological model universe interacting with perfect fluid in Lyra geometry with different cases. Banerjee et al.[131] have investigated Bianchi type-I cosmological models with viscous fluid in higher dimensional space time. Also Krori et al.[132] studied Bianchi type-I string cosmological model in higher dimensions.  Many authors [133-140] have studied Bianchi type models in order to examine the role of certain anisotropic sources during the formation of the large-scale structures that we see in the present universe and for better understanding the small amount of observed anisotropy in the universe.\\
\\
Motivated from the above, in this paper we discussed about cosmic string universes by considering Locally Rotationally Symmetric (LRS) Bianchi Type I metric in  five dimensional Lyra manifold with two different cases and find out the realistic solutions which are supporting to the present observational facts. We also found that the geometry itself of Lyra manifold behaves as a new source of dark energy which will be beneficial for further research work.\\
\\
This paper is organized as follows: In  section 2 we are presenting the field equations and their solutions. In Section 3, Physical interpretations of the solutions are presented. In the last section, conclusion of our work is presented.\\

\section{Field Equations and their Solutions  }
\label{sect:Obs}
The Einstein's field equations based on Lyra's manifold can be written as
\begin{equation}
R_{ij}-\frac{1}{2}g_{ij}R+\frac{3}{2}\phi_{i}\phi_{j}-\frac{3}{4}g_{ij}\phi_k\phi^k=-{\chi}T_{ij}
\end{equation}
where $\phi_{i}$ is the displacement vector and other symbols have their usual meanings in Riemannian geometry.
\\
\\
Let us consider the five dimensional LRS-Bianchi type - I metric in the form
\begin{equation}
ds^{2}=A^{2}dx^{2}+B^{2}dy^{2}+B^{2}dz^{2}+C^{2}d\psi^{2}-dt^{2}
\end{equation}
where A, B and C are functions of cosmic time $"t"$ only.\\
\\
The displacement vector $\phi_{i}$ is defined as
\begin{equation}
\phi_{k}=(0,0,0,0,\beta(t))
\end{equation}
and the energy momentum tensor for cosmic strings is taken as
\begin{equation}
T_{ij}=-\rho U_{i}U_{j}-\lambda X_{i}X_{j}
\end{equation}
where, $\rho=\rho_{p}+\lambda$ , is the energy density of the cloud of strings with particles attached to them, $\rho_{p}$ being the rest energy density of particles attached to the strings and $\lambda$ is the tension density of the strings.\\
 \\
 The five velocity vector $U^{i}$ of the cloud of particles and the direction of string $X^{i}$ are given by
 \begin{equation}
U^{i}=(0,0,0,0,1) ~~ and ~~  X^{i}=(A^{-1},0,0,0,0)
\end{equation}
 The direction of strings satisfies
\begin{equation}
U_{i}U^{i}=-X_{i}X^{i}=-1~~ and ~~U_{i}X^{i}=0
\end{equation}\\
Using the co - moving coordinate system, the field equation (1) gives the following equations
\begin{equation}
-\left(\frac{\dot{B}}{B}\right)^2-2\frac{\dot{A}\dot{B}}{AB}-2\frac{\dot{B}\dot{C}}{BC}-2\frac{\dot{A}\dot{C}}{AC}+\frac{3}{4}\beta^2=-\chi\rho
\end{equation}
\begin{equation}
2\frac{\ddot{B}}{B}+\frac{\ddot{C}}{C}+\left(\frac{\dot{B}}{B}\right)^2+2\frac{\dot{B}\dot{C}}{BC}+\frac{3}{4}\beta^2=\chi\lambda
\end{equation}
\begin{equation}
\frac{\ddot{A}}{A}+\frac{\ddot{B}}{B}+\frac{\ddot{C}}{C}+\frac{\dot{A}\dot{B}}{AB}+\frac{\dot{B}\dot{C}}{BC}+\frac{\dot{A}\dot{C}}{AC}+\frac{3}{4}\beta^2=0
\end{equation}
\begin{equation}
\frac{\ddot{A}}{A}+2\frac{\ddot{B}}{B}+\left(\frac{\dot{B}}{B}\right)^2+2\frac{\dot{A}\dot{B}}{AB}+\frac{3}{4}\beta^2=0
\end{equation}
where dot denotes differentiation with respect to $"t"$.\\
\\
Now subtracting (9) from (10) we have
\begin{equation}
\frac{\ddot{B}}{B}-\frac{\ddot{C}}{C}+\left(\frac{\dot{B}}{B}\right)^2+\frac{\dot{A}\dot{B}}{AB}-\frac{\dot{B}\dot{C}}{BC}-\frac{\dot{A}\dot{C}}{AC}=0
\end{equation}
\\
\\
\textbf{Case - I :}
A set of solution of the equations (7)-(10) are
\begin{equation}
A=a_0e^{2b_{0}t}[b_0c_0t^2+(b_0c_1+2c_0)t+b_0c_2+c_1]^{-1}
\end{equation}
\begin{equation}
B=e^{-b_{0}t+b_1}
\end{equation}
\begin{equation}
C=c_0t^2+c_1t+c_2
\end{equation}
where $a_{0} , b_{0} , c_{0} , c_{1}$ and $c_{2}$ are constants.\\
\\
Then from equation (10) we have

\[\begin{split}
\beta^2&=\frac{4}{3}[4b_0^3c_0^2t^3+6b_0^3c_0c_1t^2+6b_0^2c_0^2t^2+4b_0^3c_0c_2t+2b_0^3c_1^2t+6b_0^2c_0c_1t-4b_0c_0^2t+2b_0^3c_1c_2\\
                  &+6b_0^2c_0c_2-2b_0c_0c_1-8c_0^2]\times\left[b_0c_0t^2+b_0c_1t+2c_0t+b_0c_2+c_1\right]^{-2}-4b_0^2
\end{split}\]
\\
i.e.
\\
\begin{equation}
\beta^2=\frac{4}{3}\frac{\beta_{1}t^3+\beta_{2}t^2+\beta_{3}t+\beta_{4}}{[b_0c_0t^2+b_0c_1t+2c_0t+b_0c_2+c_1]^{2}}-4b_0^2
\end{equation}
\\
\\
where $\beta_{1}=4b_0^3c_0^2$ , $\beta_{2}=6b_0^3c_0c_1+6b_0^2c_0^2$ , $\beta_{3}=4b_0^3c_0c_2+2b_0^3c_1^2+6b_0^2c_0c_1-4b_0c_0^2$ and $\beta_{4}=2b_0^3c_1c_2+6b_0^2c_0c_2-2b_0c_0c_1-8c_0^2$ are constants.
\\
\\
Again equation (7) gives

\begin{equation}
\begin{split}
\chi\rho&=[2b_0^2c_0^3t^4+4b_0^2c_0^2c_1t^3+3b_0^2c_0c_1^2t^2+b_0^2c_1^3t+8b_0c_0^2c_2t-2b_0c_0c_1^2t+b_0^2c_1^2c_2\\
        &-2b_0^2c_0c_2^2+4b_0c_0c_1c_2-b_0c_1^3+8c_0^2c_2-2c_0c_1^2]\times\left(c_0t^2+c_1t+c_2\right)^{-1}\\
        &\times\left[b_0c_0t^2+b_0c_1t+2c_0t+b_0c_2+c_1\right]^{-2}
\end{split}
\end{equation}
\\
Also from equation (8) we have

\begin{equation}
\begin{split}
\chi\lambda&=-[8b_0^2c_0^3t^4+10b_0^2c_0^2c_1t^3+12b_0c_0^3t^3+12b_0^2c_0c_1^2t^2+18b_0c_0^2c_1t^2+4b_0^2c_1^3t+\\
           &10b_0c_0c_1^2t-4b_0c_0^2c_2t+4b_0^2c_1^2c_2-8b_0^2c_0c_2^2+2b_0c_1^3-2b_0c_0c_1c_2-2c_0c_1^2+8c_0^2c_2]\\
           &\times\left(c_0t^2+c_1t+c_2\right)^{-1}\times\left[b_0c_0t^2+b_0c_1t+2c_0t+b_0c_2+c_1\right]^{-2}
\end{split}
\end{equation}
\\
Here the expansion factor $\theta$ is given by

\begin{equation}
\theta = \frac{2c_0^{2}t^2+2c_0c_1t+c_1^{2}-2c_0c_2}{[b_0c_0t^2+(b_0c_1+2c_0)t+b_0c_2+c_1](c_0t^2+c_1t+c_2)}
\end{equation}
\\
And the volume $V$ is obtained as

\begin{equation}
V=\frac{k(c_0t^2+c_1t+c_2)}{b_0c_0t^2+(b_0c_1+2c_0)t+b_0c_2+c_1}
\end{equation}
\\
In this case we get the shear scalar $\sigma$ in the form

\begin{equation}
\begin{split}
\sigma&=\frac{1}{\sqrt{18}}[(k_{11}t^{4}+k_{12}t^{3}+k_{13}t^{2}+k_{14}t+k_{15})^{2}\\
   &+2(k_{21}t^{4}+k_{22}t^{3}+k_{23}t^{2}+k_{24}t+k_{25})^{2}\\
   &+(k_{31}t^{4}+k_{32}t^{3}+k_{33}t^{2}+k_{34}t+k_{35})^{\frac{1}{2}}]\\
   &\times[b_0c_0t^2+(b_0c_1+2c_0)t+b_0c_2+c_1]^{-1}\times(c_0t^2+c_1t+c_2)^{-1}
\end{split}
\end{equation}
\\
\\
where $k_{ij}$'s ; $i=1,2,3$ and $j=1,2,3,4,5$ are constants.
\\
\\
Also the deceleration parameter $q$ is given by

\[\begin{split}
q&=3[4b_{0}c_{0}^{4}t^{5}+10b_{0}c_{0}^{3}c_{1}t^{4}+4c_{0}^{4}t^{4}-8b_{0}c_{0}^{3}c_{2}t^{3}+12b_{0}c_{0}^{2}c_{1}^{2}t^{3}+8c_{0}^{3}c_{1}t^{3}\\
   &-12b_{0}c_{0}^{2}c_{1}c_{2}t^{2}+8b_{0}c_{0}c_{1}^{3}t^{2}-16c_{0}^{3}c_{2}t^{2}+10c_{0}^{2}c_{1}^{2}t^{2}-12b_{0}c_{0}^{2}c_{2}^{2}t+2b_{0}c_{1}^{4}t\\
   &-16c_{0}^{2}c_{1}c_{2}t+6c_{0}c_{1}^{3}t-6b_{0}c_{0}c_{1}c_{2}^{2}+2b_{0}c_{1}^{3}c_{2}-4c_{0}^{2}c_{2}^{2}-2c_{0}c_{1}^{2}c_{2}+c_{1}^{4}]\\
   &\times\left[2c_{0}^{2}t^{2}+2c_{0}c_{1}t+c_{1}^{2}-2c_{0}c_{2}\right]^{-2}-1
\end{split}\]
\\
i.e.
\\
\begin{equation}
q=\frac{3(a_{1}t^{5}+a_{2}t^{4}+a_{3}t^{3}+a_{4}t^{2}+a_{5}t+a_{6})}{(2c_{0}^{2}t^{2}+2c_{0}c_{1}t+c_{1}^{2}-2c_{0}c_{2})^{2}}-1
\end{equation}
\\
\\
where, \\
$a_{1}=-4b_{0}c_{0}^{4}$ \\
$a_{2}=10b_{0}c_{0}^{3}c_{1}+4c_{0}^{4}$ \\
$a_{3}=-8b_{0}c_{0}^{3}c_{2}+12b_{0}c_{0}^{2}c_{1}^{2}+8c_{0}^{3}c_{1}$ \\ $a_{4}=-12b_{0}c_{0}^{2}c_{1}c_{2}+8b_{0}c_{0}c_{1}^{3}-16c_{0}^{3}c_{2}+10c_{0}^{2}c_{1}^{2}$ \\
$a_{5}=-12b_{0}c_{0}^{2}c_{2}^{2}+2b_{0}c_{1}^{4}-16c_{0}^{2}c_{1}c_{2}+6c_{0}c_{1}^{3}$    $and$\\
$a_{6}=-6b_{0}c_{0}c_{1}c_{2}^{2}+2b_{0}c_{1}^{3}c_{2}-4c_{0}^{2}c_{2}^{2}-2c_{0}c_{1}^{2}c_{2}+c_{1}^{4}$ \\
are constants.
\\
\\
Also from (16) and (17) we have

\begin{equation}
\begin{split}
\chi\rho-\chi\lambda&=[10b_0^2c_0^3t^4+14b_0^2c_0^2c_1t^3-12b_0c_0^3t^3-9b_0^2c_0c_1^2t^2-18b_0c_0^2c_1t^2+4b_0c_0^2c_2t-\\
                    &10b_0c_0c_1^2t-4b_0^2c_1^3t+6b_0^2c_0c_2^2+6b_0c_0c_1c_2-3b_0^2c_1^2c_2-3b_0c_1^3]\\
                    &\times\left(c_0t^2+c_1t+c_2\right)^{-1}\times\left[b_0c_0t^2+b_0c_1t+2c_0t+b_0c_2+c_1\right]^{-2}
\end{split}
\end{equation}
\\
By taking particular values of the integrating constants  as  $a_{0}=b_{0}=c_{0}=c_{1}=c_{2}=1$, the variations of some parameters with respect to time of the \textbf{Case-I} are shown in \textbf{Figures 1-4}.
\\
\\
\includegraphics{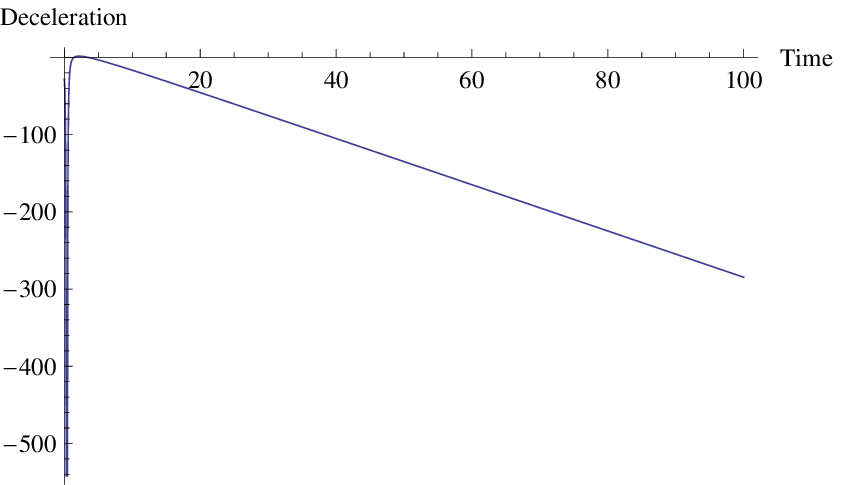}
\\
\\
\textbf{Figure-1 : The Variation of Deceleration Parameter $q$ vs. Time $t$. Here $a_{0}=b_{0}=c_{0}=c_{1}=c_{2}=1$}\\
\\
\\
\includegraphics{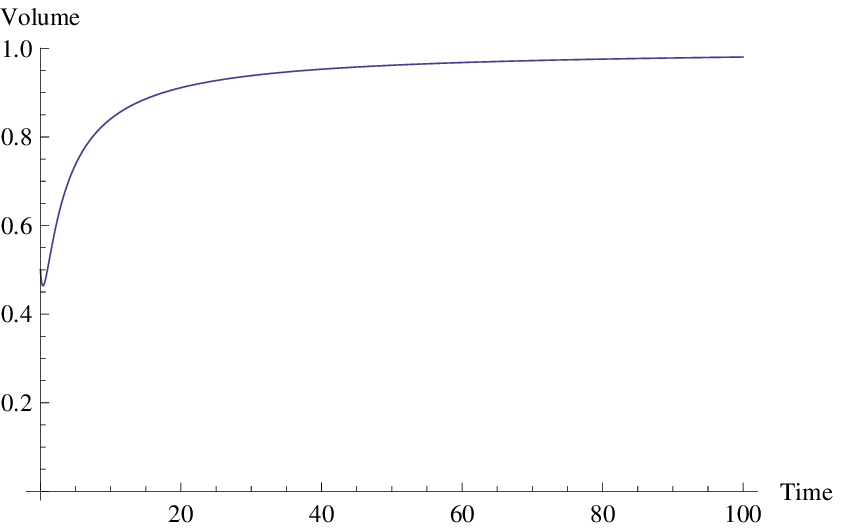}
\\
\\
\textbf{Figure-2 : The Variation of Volume $V$ vs. Time $t$. Here $a_{0}=b_{0}=c_{0}=c_{1}=c_{2}=1$}\\
\\
\\
\includegraphics{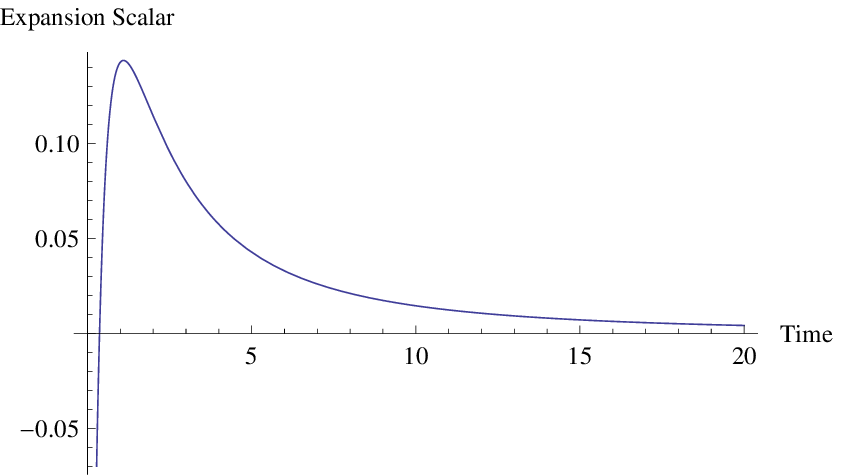}
\\
\\
\textbf{Figure-3 : The Variation of Expansion Scalar $\theta$ vs. Time $t$. Here $a_{0}=b_{0}=c_{0}=c_{1}=c_{2}=1$}\\
\\
\\
\includegraphics{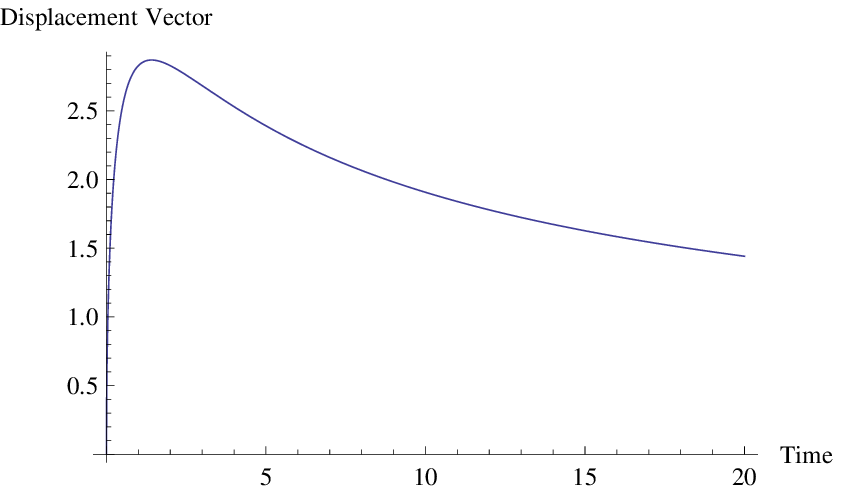}
\\
\\
\textbf{Figure-4 : The Variation of Displacement Vector $\beta$ vs. Time $t$. Here $a_{0}=b_{0}=c_{0}=c_{1}=c_{2}=1$}\\
\\
\\
\textbf{Case - II} :
Again

\begin{equation}
A=e^{ab_0t+ab_1}
\end{equation}
\begin{equation}
B=e^{b_0t+b_1}
\end{equation}
and
\begin{equation}
C=c_1e^{b_0t}+c_2e^{-(2+a)b_0t}
\end{equation}
are another set of solutions of the equations (7) to (10), where $a ,b_{0} , b_{1} , c_{1}$ and $c_{2}=1$ are constants.\\
\\
Now using these values of $A$, $B$ and $C$ we get from (7), (8) and (10)

\begin{equation}
\chi\rho=a^2+2a+3+(2a+1)b_0^2+2(a+1)b_0\frac{b_0c_1e^{b_0t}-(2+a)b_0c_2e^{-(2+a)b_0t}}{c_1e^{b_0t}+c_2e^{-(2+a)b_0t}}
\end{equation}

\begin{equation}
\chi\lambda=(3b_0^2-a^2-2a-3)+\frac{3b_0^2c_1e^{b_0t}+(2+a)ab_0^2c_2e^{-(2+a)b_0t}}{c_1e^{b_0t}+c_2e^{-(2+a)b_0t}}
\end{equation}
\\
and
\begin{equation}
\beta^2=-\frac{4}{3}(a^2+2a+3)
\end{equation}
\\
In this case we have
Expansion factor
\begin{equation}
\theta=(a+3)b_0c_1e^{b_0t}[c_1e^{b_0t}+c_2e^{-(2+a)b_0t}]
\end{equation}
\\
Deceleration parameter
\begin{equation}
q=-[1+3\frac{c_2}{c_1}e^{-(3+a)b_{0}}]
\end{equation}
\\
and the volume
\begin{equation}
V=[c_1e^{b_0t}+c_2e^{-(2+a)b_0t}]e^{(a+2)b_0t+(a+1)b_1}
\end{equation}
\\
And the shear scalar $\sigma$ is given by

\begin{equation}
\sigma^{2}=\frac{1}{\sqrt{18}}[9(a^{2}+2)b_{0}^{2}+\frac{9[b_0c_1e^{b_0t}-(2+a)b_0c_2e^{-(2+a)b_0t}]}{c_1e^{b_0t}+c_2e^{-(2+a)b_0t}}-\frac{2(a+3)b_0c_1^{2}e^{b_0t}}{[c_1e^{b_0t}+c_2e^{-(2+a)b_0t}]^{2}}]^{\frac{1}{2}}
\end{equation}
\\
In this case,

\begin{equation}
\chi\rho-\chi\lambda=(4+2a)b_0^{2}+\frac{(5+2a)b_0^{2}c_1e^{b_0t}-(a^2+4a+4)b_0^{2}c_2e^{-(2+a)b_0t}}{c_1e^{b_0t}+c_2e^{-(2+a)b_0t}}
\end{equation}
\\
The variations of some parameters with respect to time of the \textbf{Case-II}  are shown in \textbf{Figures 5-7}, by considering particular values of the integrating constants  as  $a=b_{0}=b_{1}=c_{1}=c_{2}=1$.
\\
\\
\includegraphics{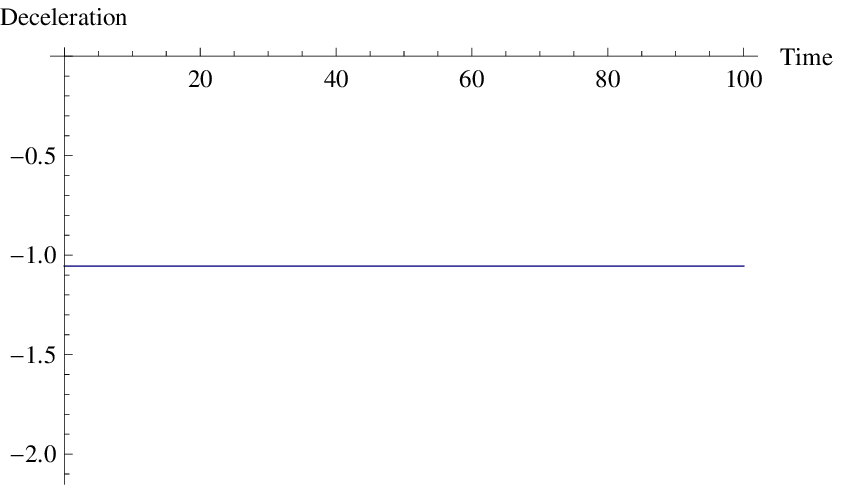}
\\
\\
\textbf{Figure-5 : The Variation of Deceleration Parameter $q$ vs. Time $t$. Here $a=b_{0}=b_{1}=c_{1}=c_{2}=1$}\\
\\
\\
\includegraphics{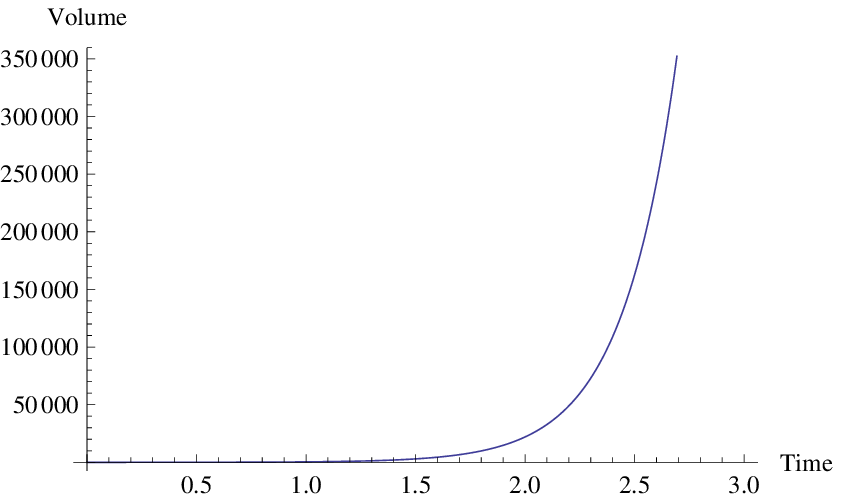}
\\
\\
\textbf{Figure-6 : The Variation of Volume $V$ vs. Time $t$. Here $a=b_{0}=b_{1}=c_{1}=c_{2}=1$}\\
\\
\\
\includegraphics{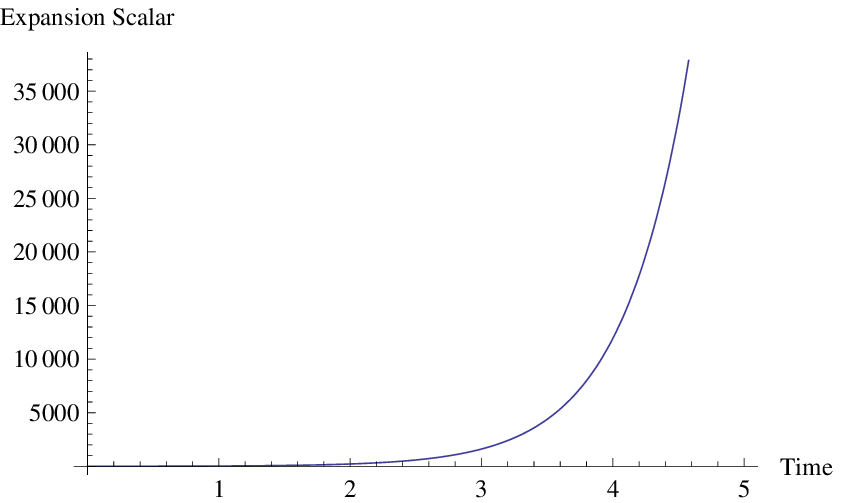}
\\
\\
\textbf{Figure-7 : The Variation of Expansion Scalar $\theta$ vs. Time $t$. Here $a=b_{0}=b_{1}=c_{1}=c_{2}=1$}\\
\\
\\

\section{Physical Interpretations of the solutions}
\label{sect:Interpretation}

Here in a five-dimensional Lyra manifold, we have constructed a model universe containing string cloud with standard matter that interacts with the vector field $\phi_{k}$. With the assumption of a constant displacement vector field, our model behaves as a dark energy model universe, particularly the phantom energy type. The displacement field, which is considered as a component of the total energy, plays the role of dark energy. In one case our universe seems to be dominated by cosmological constant type of dark energy.\\
\\
In \textbf{Case - I} we see that the energy density of our universe has a finite value at $t=0$, that is, at the beginning of the evolution; and even when $t\rightarrow\infty$ it has some finite value and there is no singularity; thereby there is possibility that our model is that of the oscillating type. The string density has some definite value at the beginning of the epoch until it gradually decreases so that it is zero at time $t$ given by

\[(b_0c_0t^2+b_0c_1t+2c_0t+b_0c_2+c_1)^{2} = 4b_{0}^{4}c_{0}^{2}t^{4}+8b_{0}^{4}c_{0}c_{1}t^{3}\]
\\
From the expression of the deceleration parameter and its variation with time as shown in the above \textbf{Fig-1} we see that our universe is undergoing an accelerated expansion. Again, the values of the volume $V$ and the expansion factor $\theta$ show that our universe is expanding without any singularity, though at $t=0$ and $t\rightarrow\infty$ the expansion rate is constant.\\
\\
Thus we see that, with the advent of time, the density of string decreases more rapidly than density of the particles attached to them indicating that our universe becomes a universe consisting only of particles, matter and energy, where strings become invisible.\\
\\
In this model the gauge function $\beta^{2}$ is found to be constant at the beginning of the evolution and gradually decreases with time until it takes a finite value (another constant value) at infinite time. Here it is seen that the displacement vector, interacting with pressureless matter, plays the same role as the cosmological constant. From the above characteristics we may conclude that this model universe is one containing and dominated by dark energy.\\
\\
In \textbf{Case - II} the energy density of our universe as well as the tension density of the string are found to be decreasing functions of time, and the string density becomes zero at the point $6b_{0}^{2}=a^{2}+2a+3$. And their difference takes a definite quantity as $t\rightarrow\infty$ which implies that as time passes the string vanish leaving the particles in this universe. The volume of this universe is found here to expand at a high rate. And more from the expressions of the expansion factor and the deceleration parameter with their variations with time as shown in the above graphical representations of this case  it can be concluded that the universe is expanding at an accelerated rate. Since the energy density was negative in the past, increases towards a positive value then decreases toward zero with time. The expansion of the early universe is due to the negative energy density with negative pressure and the late-time accelerated expansion comes from the positive energy and negative pressure which behaves like a dark energy. It is amazing that the non-singular universe is expanding with time whereas a transition from negative energy density to positive energy density occurs its dynamical evolution. Similar scenario occurs in Kim and Yoon [141] and negative energy density in accelerated universe was discussed recently by Sawicki and Vikman [142]. Thus here it is seen that our model behaves as a dark energy filled universe.\\
\\
In a phantom dark energy filled universe, for the phantom scalar $\phi_{ph}$ we know that $(\dot{\phi}_{ph})^{2}<0$. Strikingly similar in our problem here it is found that the displacement vector $\beta^{2}$ is less than zero. Thus it is interesting here to think, without loss of generality, of the Lyra manifold defining itself as a source of dark energy, as our universe here behaves and have the characteristics of a universe filled with phantom form of dark energy. Thus it is very much stimulating to see that a universe in Lyra manifold can define itself as a dark energy model universe, perhaps the dark energy here being the phantom energy type in most of the cases. Thus the geometry itself of the Lyra manifold may be taken as a hidden source of dark energy. Large number of literatures exists to address the observational fact of the current expansion and evolution of universe from the higher dimensional point of view as we mentioned above. But regarding the source of dark energy and its connections to extra-dimensions, we are not aware of models similar to the one developed in this paper. Further details, consequences and numerical confrontations with observations are in progress.

\section{Conclusion}

 Different mines provide different opinions about our universe but as a concluding remarks, we can say that our model behaves as a dark energy model universe, particularly the phantom energy type. The displacement field, which is considered as a component of the total energy, plays the role of dark energy. And in one case our universe seems to be dominated by cosmological constant type of dark energy. However what we have presented is nothing else but a toy model which has to be worked out in detail, first of all by matching the observations. It will be the issue of our forthcoming paper.\\

\textbf{Aacknowledgements:} We would like to thank Prof H.Saller, Editor in Chief, IJTP and anonymous referee for their cooperation and valuable suggestions that helped us to significantly improve our manuscript to this level.



\end{document}